\documentclass[twocolumn]{aastex63}
\usepackage{amsmath,amssymb,amstext}
\usepackage{gensymb}

\shorttitle{Multi-Epoch Terrestrial Transmission Spectra}
\shortauthors{May, Taylor, Komacek, et al.}

\newcommand\jwst{{\em JWST}}
\graphicspath{{./}{plots/}}

\begin{document}
\title{Water Ice Cloud Variability \& Multi-Epoch Transmission Spectra of TRAPPIST-1e}

\correspondingauthor{E. M. May}
\email{Erin.May@jhuapl.edu}

\author[0000-0002-2739-1465]{E. M. May}
\affiliation{Johns Hopkins APL, 11100 Johns Hopkins Rd, Laurel, MD 20723, USA}

\author[0000-0003-4844-9838]{J. Taylor}
\affiliation{Department of Physics (Atmospheric, Oceanic and Planetary Physics), University of Oxford, Parks Rd, Oxford, OX1 3PU, UK}

\author[0000-0002-9258-5311]{T. D. Komacek}
\affiliation{Department of the Geophysical Sciences, University of Chicago, Chicago, IL 60637, USA}

\author[0000-0002-2338-476X]{M. R. Line}
\affiliation{School of Earth \& Space Exploration, Arizona State University, Tempe, AZ 85257, USA}

\author[0000-0001-9521-6258]{V. Parmentier}
\affiliation{Department of Physics (Atmospheric, Oceanic and Planetary Physics), University of Oxford, Parks Rd, Oxford, OX1 3PU, UK}

\begin{abstract}
The precise characterization of terrestrial atmospheres with the James Webb Space Telescope (\jwst) is one of the utmost goals of exoplanet astronomy in the next decade. With \jwst's impending launch, it is crucial we are well prepared to understand the subtleties of terrestrial atmospheres - particularly ones we may have not needed to consider before due to instrumentation limitations. In this work we show that patchy ice cloud variability is present in the upper atmospheres of M-dwarf terrestrial planets, particularly along the limbs. Here we test whether these variable clouds will introduce unexpected biases in the multi-epoch observations necessary to constrain atmospheric abundances. Using 3D {\tt{ExoCAM}} general circulation models (GCMs) of TRAPPIST-1e, we simulate five different climates with varying pCO$_2$ to explore the strength of this variability. These models are post-processed using NASA Goddard's Planetary Spectrum Generator (PSG) and {\tt{PandExo}} to generate simulated observations with \jwst's NIRSpec PRISM mode at 365 different temporal outputs from each climate. Assuming the need for 10 transits of TRAPPIST-1e to detect molecular features at great confidence, we then use {\tt{CHIMERA}} to retrieve on several randomly selected weighted averages of our simulated observations to explore the effect of multi-epoch observations with variable cloud cover along the limb on retrieved abundances. We find that the variable spectra do not affect retrieved abundances at detectable levels for our sample of TRAPPIST-1e models.
\end{abstract}

\section{Introduction}
\par With the coming launch of the James Webb Space Telescope (\jwst) we will gain unprecedented ability to characterize the atmospheres of terrestrial exoplanets. With this exciting observational advancement comes a renewed focus on model predictions for sub-Jovian atmospheres 
(e.g., \citealp{Yang:2013,Kaspi2015,Koll2016,Turbet:2016aa,Fujii:2017aa,Kang:2019ab,May2020b}, see \citealp{Pier2019,Shields:2019aa} for recent reviews). Importantly, all of these works focus on time invariable phenomena and to date have largely considered only the time-averaged planetary climate. Notably, the atmospheres of hot gas giants may be time-variable at an observationally detectable level  \citep{Rauscher:2007aa,Komacek:2020ac,Menou:2020aa}. As a result, it is important to consider these time-dependent effects when moving to characterize a new class of planets. 
\par Clouds play an important role in atmospheric circulation, providing radiative feedback that both cools the atmosphere by blocking additional stellar radiation from penetrating deep into the atmosphere, while also warming the lower atmosphere due to the greenhouse effect of cloud decks. The temperature gradients that arise from this spatially variable radiative forcing can drive winds and shape the global circulation. While we know them to be a key player in shaping the observable properties of exoplanet atmospheres \citep[e.g.,][]{Roman2020,Parmentier2021}, their complexity often results in clouds being ‘under-modeled’ compared to the rest of the atmosphere. Because of the important feedback roles of clouds, it is imperative that we use accurate (within computational limits) treatments of clouds - especially to the extent that their 3D nature, radiative feedback, and variability (formation and dissipation) will have a non-negligible impact on observed properties. Recent work by \cite{Charnay:2020aa} presented evidence for cloud variability on sub-Neptunes in 3D climate models, specifically for K2-18b, but did not study how this variability may affect observations.
\par To date, characterizing the atmospheres of terrestrial planets has been difficult due to their higher metallicity atmospheres, corresponding to small atmospheric scale heights (H $\propto$ 1/$\mu$) and transmission signals ($\propto$ HR$_p$/R$_s^2$). For that reason, primary targets for \jwst{ }observations of terrestrial planets are around M-dwarf hosts due to the optimal size ratios of the star and planet. Among the GTO transiting exoplanet programs, 4 transits of TRAPPIST-1e and 2 transits of TRAPPIST-1d are planned with NIRSpec PRISM (Program ID 1331, PI: Nikole Lewis and Program ID 1201, PI: David Lafreni\`{e}re; respectively) and 5 transits of TRAPPIST-1f are planned with NIRISS SOSS (Program ID 1201, PI: David Lafreni\`{e}re). In addition to the planned GTO and ERS programs, there is a desire in the community to spend significant \jwst\ GO time detecting atmospheres around the habitable zone TRAPPIST-1 planets \citep{Gillon2020}. Typical feature sizes for a planet similar to TRAPPIST-1e with an entirely clear atmosphere are of order 50 ppm, and with an estimated noise floor of approximately $\sim$10 ppm \citep{Schlawin2020}, \jwst\ will be capable of detecting molecular absorption in the atmospheres of cloud-free small planets. As planets in the TRAPPIST-1 system are likely to be tidally locked (however, see \citealp{Leconte:2015}), dayside convection may lead to copious cloud formation that impacts the detectability of molecular features in the atmospheres of these planets in transmission \citep{Fauchez2019,Komacek:2020aa,Suissa:2020aa}.

\par The broad wavelength range of NIRSpec PRISM makes it the most effective instrument for terrestrial planet observations due to the additional wavelength coverage, even preferred over higher precision observations with other modes \citep{Greene2016,Batalha&Line2017}. In particular, observations at the short wavelengths accessible with NIRSpec PRISM allow for breaking the degeneracy between mean molecular weight and patchy cloud coverage for large planets \citep{Line2016}. \cite{KrissansenTotton2018} find that 10 transits with NIRSpec PRISM should be sufficient to detect the biosignature combination of CH$_4$ and CO$_2$ and lack of  CO in cloud-free conditions. \cite{LustigYaeger2019} further predict that for a CO$_2$ dominated atmosphere, ~$\sim$ 7-8 transits of TRAPPIST-1e are needed with NIRSpec PRISM to detect an atmosphere at 5$\sigma$, with only Venus-like and H$_2$O dominated atmospheres requiring more than 10 transits for such a detection. In addition, \cite{Fauchez2019} find that the 4.3 $\micron$ CO$_2$ line in a modern and Archean Earth-like atmosphere, and a CO$_2$ dominated atmosphere, are detectable at 3$\sigma$ with NIRSpec PRISM in less than 15 transits, even in the presence of (unchanging) clouds and hazes. However, the above predictions for TRAPPIST-1e require that the atmospheric cloud cover is constant and unchanging -- an assumption that we seek to test here for TRAPPIST-1e.
\par We first present 3D models of TRAPPIST-1e for a range of atmospheric compositions (simulated with five different CO$_2$ partial pressures, pCO$_2$) and study the impact that the pronounced variability of ice clouds in the upper atmosphere has on simulated observations. Using the Planetary Spectrum Generator \citep[PSG,][]{Villanueva2018} and PandExo \citep{Batalha2017}, we generate limb-averaged synthetic NIRSpec PRISM observations for every temporal output from the last Earth year of model time from our five GCMs. For each atmospheric case, we generate 10 random combinations of 10 synthetic spectra to simulate the impact that multi-epoch observations will have on future observations of TRAPPIST-1e. We then use CHIMERA \citep{Line2013a,Line2013b, Tremblay:2020} to retrieve atmospheric abundances for each of our random combinations to compare to input values. Through this, we are able to determine the impact of variability on proposed multi-epoch observations of TRAPPIST-1e with \jwst's NIRSpec PRISM mode. 
\par In Section \ref{GCMs} we overview our 3D General Circulation Models (GCMs). Section \ref{PSG} discusses our model post-processing using the Planetary Spectrum Generator. In Section \ref{PandExo} we describe our simulated \jwst{ }observations of the modeled planets using PandExo. Section \ref{Retrievals} describes the atmospheric retrievals performed on the simulated data which we compare to our input compositions in Section \ref{Results}. Finally, in Section \ref{Conclusions} we present the conclusions of this work.

\begin{figure*}[ht!]
    \centering
    \includegraphics[width = 1\textwidth]{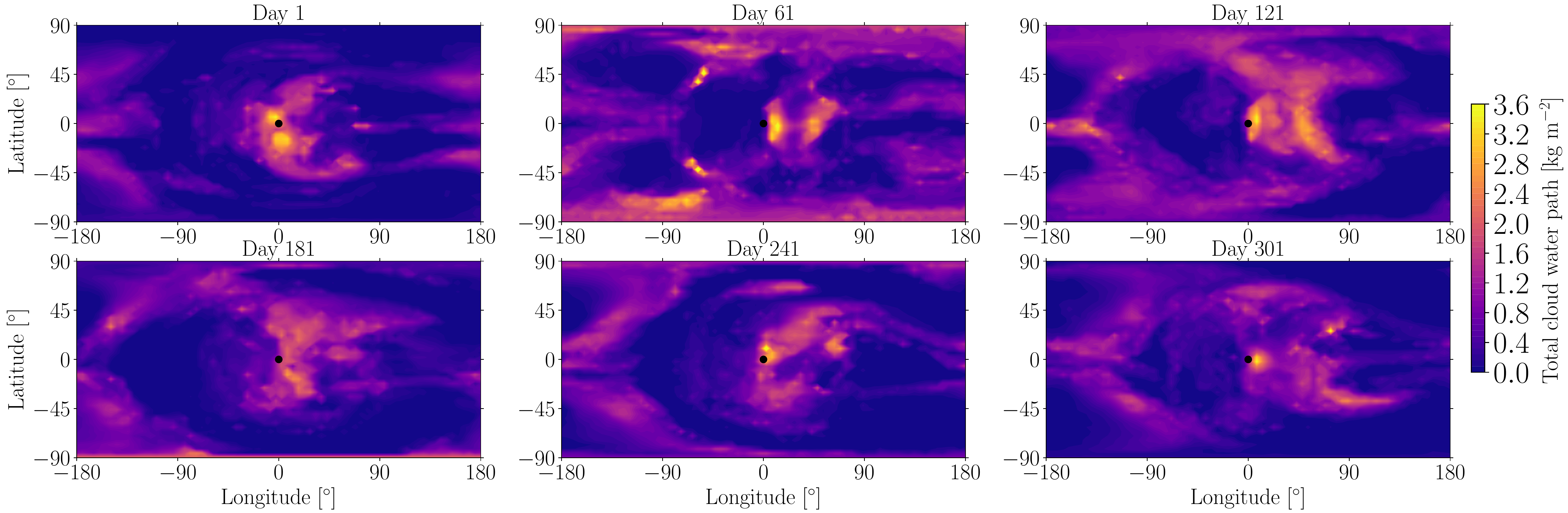}
    \caption{Maps of the column-integrated cloud water path at six different days (each separated by 60 days) from the GCM simulation with 100 mbar of CO$_2$. The atmospheric cloud mass is greatly time-variable, with most locations experiencing both nearly cloud-free days and days with strong cloud coverage. As a result, the total cloud water path at the terminator strongly varies with time, though regions near the substellar point are perpetually cloudy.}
    \label{fig:mapscloudmass}
\end{figure*}

\section{3D General Circulation Models} \label{GCMs}
\subsection{Model setup}
In this work, we apply the {\tt ExoCAM}\footnote{\url{https://github.com/storyofthewolf/ExoCAM}} General Circulation Model (GCM) to study climate variability in the atmosphere of TRAPPIST-1e. {\tt ExoCAM} is a modified version of the Community Atmosphere Model (CAM) v4.0 \citep{Neale:2010aa} that includes the novel non-grey correlated-k radiative transfer scheme {\tt ExoRT}\footnote{\url{https://github.com/storyofthewolf/ExoRT}}. This enables {\tt ExoCAM} to model planets over a much broader range of climate states than Earth. {\tt ExoCAM} has been used in a wide array of previous studies of the atmospheric circulation of early Earth and exoplanets orbiting a broad range of stellar types (e.g., \citealp{kopparapu2017,Wolf:2017aa,Haqq2018,Komacek2019,Yang:2019aa,Suissa:2020aa,Wei:2020aa}).

Similar to \cite{Wolf:2017aa}, we perform a grid of GCMs varying the atmospheric CO$_2$ partial pressure. Specifically, we conduct {\tt ExoCAM} simulations for CO$_2$ partial pressures of $10^{-4}, 10^{-3}, 10^{-2}, 10^{-1},$ and $1~\mathrm{bar}$, encompassing most of the range of climate states considered by \cite{Wolf:2017aa}. We assume an aquaplanet with plentiful surface water, and include 1 bar of N$_2$ in all simulations. As a result, our model atmospheres are composed purely of 1 bar of N$_2$, CO$_2$ at the assumed partial pressure, and H$_2$O determined by its saturation vapor pressure. As varying CO$_2$ alone allows us to cover a broad range of both cold and hot climate states, we do not include other greenhouse gases (e.g., CH$_4$) or additional atmospheric constituents (e.g., O$_2$/O$_3$) in order to conduct a clean parameter sweep with a single varying parameter, pCO$_2$. Note that our assumptions for model atmospheric composition imply that the total surface pressure is different in each simulation with varying partial pressures of CO$_2$ and H$_2$O. The maximum surface pressure in our suite of simulations is 2 bars, while  {\tt ExoRT} is valid for pressures up to 10 bars \citep{Wolf:2014aa}. The changing background pressure does affect the mean climate state, but in this work we focus on the impact of climate variability of TRAPPIST-1e on observable properties rather than the impact of mean climate. We assume a planetary radius of $0.92$ Earth radii, a surface gravity of $9.12~\mathrm{m}~\mathrm{s}^{-2}$ and an incident stellar flux of $900.85~\mathrm{W}~\mathrm{m}^{-2}$, derived from the observations of \cite{Gillon:2017aa} and \cite{Grimm:2018aa}. For simplicity, we  assume that TRAPPIST-1e is spin-synchronized of $6.10$ Earth days (set equal to the orbital period). However, note that TRAPPIST-1e may lie in a higher-order or quasi-stable spin state \citep{Leconte:2015,Vinson:2019aa}, which requires further study to determine the impact of spin state on climate variability. 

For all of the simulations presented in this work, we use an incident stellar spectrum from the models of \cite{Allard:2007aa} for an M-dwarf star with an effective temperature of $2600~\mathrm{K}$. All simulations assume that the orbit of TRAPPIST-1e has zero eccentricity and that the obliquity of TRAPPIST-1e is zero. We further assume that the surface of the planet consists of a slab (non-dynamic) ocean with a depth of $50~\mathrm{m}$. This ocean can form sea ice, but we do not consider ocean heat transport and the modeled sea ice distribution is governed purely by thermodynamics \citep{Bitz:2012aa}. We use the sub-grid parameterization for clouds developed by \cite{Rasch:1198}, assuming that liquid water clouds have an effective radius of $14~\mu\mathrm{m}$ and with the parameterized ice cloud effective radius varying from $\approx 20-200\mu\mathrm{m}$. Convection is treated with the sub-grid scheme of \cite{Zhang:1995aa}. All simulations use a horizontal resolution of $4^\circ \times 5^\circ$ with 40 vertical levels. Our dynamical time step is 30 minutes, and the radiative time step is set to be three times the dynamical time step. Simulations are run until they reach top-of-atmosphere radiative balance, which typically takes $45-50$ Earth years of model time. In the analysis that follows, we study daily averaged output from the last year of each GCM simulation. 

\subsection{Simulated variability in cloud coverage}
As in \cite{Wolf:2017aa}, the climates in our GCM simulations of TRAPPIST-1e strongly depend on the partial pressure of CO$_2$ (pCO$_2$). Simulations with low pCO$_2$ ($\lesssim 10^{-2}~\mathrm{bar}$) have cold climates with ice coverage on much of the surface, while with increasing pCO$_2$ the atmosphere transitions to temperate and then to a hot, ice-free state with pCO$_2 = 1~\mathrm{bar}$. As a result, we find that hotter climates have greater amounts of open ocean, while cold climates have an ``eyeball'' \citep{Pierrehumbert:2011aa} of open ocean near the substellar point. In concert with the increase of the saturation vapor pressure of water with temperature from the Clausius-Clapeyron relationship, this causes simulations with larger pCO$_2$ to have more humid atmospheres, resulting in an increase in atmospheric cloudiness with increasing pCO$_2$. In our GCM simulations, the yearly average terminator-averaged cloud column mass increases by over a factor of two with varying pCO$_2$ from $10^{-4}~\mathrm{bars}$ to $1~\mathrm{bar}$.  

The local cloud coverage on a given day can be significantly different from the time-average state. As an example, Figure \ref{fig:mapscloudmass} shows maps of the vertically integrated cloud water path at six different times during the year for our simulation with pCO$_2 = 10^{-1}~\mathrm{bar}$. The cloud coverage is strongly time-variable, with most latitudes and longitudes experiencing both cloud-free and  cloudy days. This variability is driven by planetary-scale waves and the resulting superrotating equatorial jet, both of which act to transport heat and moisture from the dayside to the nightside of the planet   \citep{Labonte:2020aa}. We find that regions near the substellar point are perpetually cloudy due to vigorous convective upwelling \citep{Merlis:2010,Yang:2013}. Similar to the high-resolution simulations of \cite{Sergeev:2020aa}, the cloud coverage near the substellar point is patchy and evolves in time. Notably, similar to the case of tidally locked gas giants \citep{parmentier_2013} we find that the cloud coverage near the limb experiences some of the most significant variability. The near-terminator regions experience both some of the strongest cloud coverage not at the substellar point (e.g., near the western limb on Day 61 in Figure \ref{fig:mapscloudmass}) and at times have almost completely cloud-free regions (e.g., near the western limb on Day 1). 

\begin{figure}
    \centering
    \includegraphics[width = 0.5\textwidth]{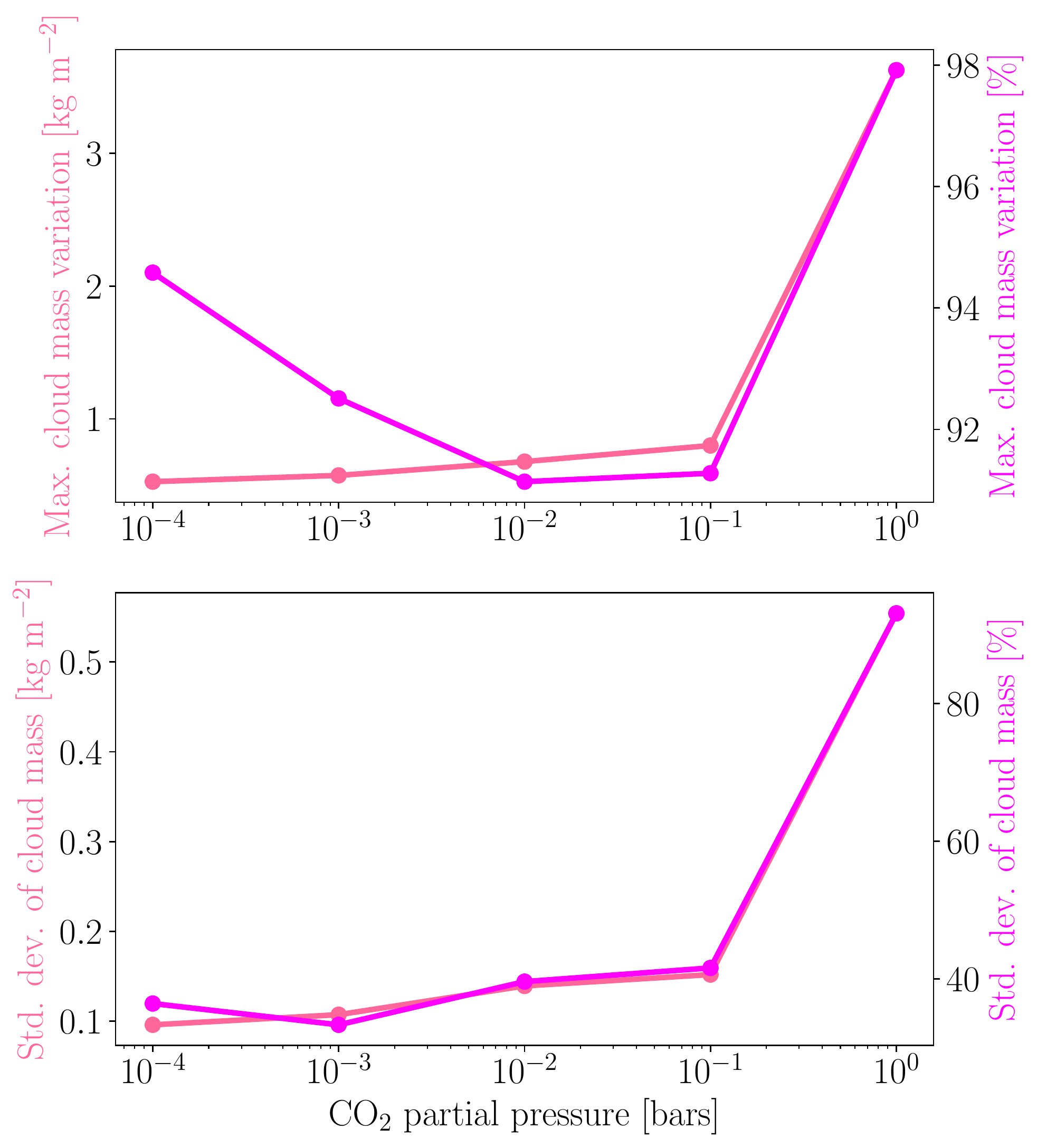}
    \caption{Top: Amplitude of the total vertical cloud water path variation at the terminator over one Earth year as a function of the CO$_2$ partial pressure. Shown are both the absolute cloud water path variation (left y-axis, orange line) and the percent cloud mass variation, normalized by the maximum cloud water path (right y-axis, magenta line). Bottom: Standard deviation of the variation in total cloud water path at the terminator over one Earth year as a function of the CO$_2$ partial pressure. Shown are both the absolute standard deviation of the variation in total cloud water path (left y-axis, orange line) and the percent standard deviation, normalized by the average total cloud water path (right y-axis, magenta line).}
    \label{fig:cloudmassvar}
\end{figure}

We find that the variability in terminator cloud coverage is large in all simulations we considered, with the maximum change in cloud coverage exceeding 90\% for all cases with varying pCO$_2$. Figure \ref{fig:cloudmassvar} shows both the absolute and relative maximum variation in terminator-averaged total cloud water path, along with the standard deviation of the distribution of daily terminator cloud water path. We find that the absolute variation and standard deviation of cloud mass both increase with increasing pCO$_2$. This is because the overall cloudiness of the atmosphere increases with pCO$_2$. Because each simulation has days where the terminator is nearly cloud free, the maximum variation and standard deviation of in cloud mass is driven by the maximum cloudiness. As a result, we expect that hotter planets (with higher pCO$_2$) will have larger-amplitude variability in their terminator cloud coverage.  

%
\begin{figure*}[ht!]
    \centering
    \includegraphics[width = \textwidth]{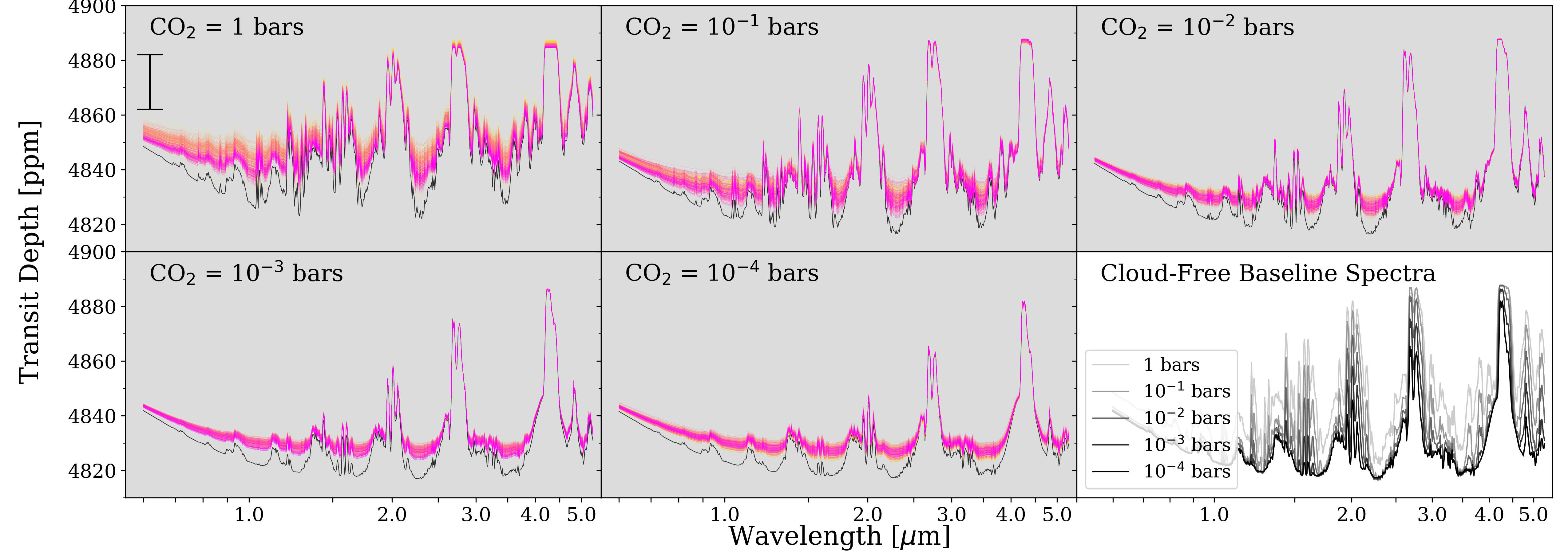}
    \caption{Transmission spectra for our five CO$_2$ partial pressures, with each panel including spectra for the last 365 Earth-days of the model (colored lines) compared to the cloud-free case for that model (averaged over 30 model days). We see clear variability in all models, with the strength of this effect decreasing with decreasing partial pressures of CO$_2$, corresponding to greater variability in the hotter atmospheres, as expected. The final (bottom right) panel in the figure includes our cloud-free baseline spectra for each CO$_2$ partial pressure case for a direct comparison. A 10 ppm error bar is shown in the first panel.}
    \label{fig:PSG}
\end{figure*}
\section{Model Post-Processing} \label{PSG}
We use the Planetary Spectrum Generator \citep[PSG,][]{Villanueva2018} to post-process our GCM outputs into transmission spectra for all resolution elements along the planet limb (longitude = +/- 90$^{\degree}$) for each day in the final Earth-year of model time. Our processed spectra include liquid water and ice clouds, the main factor driving the variability in the transmission spectra, with all input parameters (e.g., cloud particle size and mixing ratio, abundance of gaseous species) the same as those used in our GCMs. We create limb-averaged spectra for each temporal output to serve as our true model spectrum at each time output.

For comparison, we further generate cloud-free spectra from our GCMs with PSG by removing clouds ad-hoc from the GCM output. These cloud-free spectra are generated over a 30 day subset of our model output to serve as a comparison retrieval case. 

We acknowledge that by only using longitudes of +/- 90$^{\circ}$ we are not considering the full 3D effect of the light rays passing through the atmosphere from the day to night side \citep{Caldas2019}, but due to the small transit depths of the system, any differences in our limb averaging and a full 3D consideration would be within the noise of {\textit{JWST}}. We therefore present this as an initial consideration of the effect of cloud variability on observed transmission spectra, saving full 3D consideration for future work. 

In Figure \ref{fig:PSG} we show the output spectra for all five CO$_2$ partial pressure at all 365 time steps analyzed. Each panel displays a different GCM base model, while each colored line corresponds to the PSG spectrum at a different temporal output from the GCM. The black line in each figure denotes the cloud-free comparison. Liquid water and water ice clouds in the upper atmosphere drive spectral variability on scales comparable to the expected noise floor of \jwst, with the hotter atmospheres (higher partial pressures of CO$_2$) experiencing more extreme variability. The final panel shows all of our 5 cloud-free baseline spectra for all partial pressures of CO$_2$ considered for direct comparison to one another.

\section{Simulated Observations} \label{PandExo}
To explore the effects of spectral variability on multi-epoch observations of TRAPPIST-1e, we generate simulated \jwst\ observations using {\tt{PandExo}} \citep{Batalha2017}. While \cite{Batalha&Line2017} find that the combination of NIRISS SOSS + NIRSpec G395 provide the highest information content, they do not directly explore NIRSpec PRISM due to the faint magnitude limits but conclude that broad wavelength coverage is preferred over higher precision and that NIRSpec PRISM is a suitable alternative for faint host stars such as TRAPPIST-1. Further, we see significant spectral variability in the short wavelength scattering slope which is best studied with the PRISM mode or the 2$^{nd}$ order of NIRISS SOSS. Because the 2$^{nd}$ order of NIRISS SOSS requires a different observational optimization than that of the 1$^{st}$ order, the entire wavelength range is best reached with two separate observations using this instrument mode. For these reasons, we choose to model NIRSpec PRISM observations covering 0.6 to 5.3 $\micron$ to optimize our ability to constrain atmospheric constituents in retrievals on our simulated observations.

Following expected best practices for time series observations with \jwst, we assume equal times out-of-transit and in-transit, plus an extra one hour pre-transit baseline to address the tight scheduling constraints, as well as an additional half-hour pre-transit baseline for any ramp like effects. This corresponds to a total observing time of 2$\times$T$_{\mathrm{dur}}$ + 1.5 hours. We set a saturation limit of 80\% and apply TRAPPIST-1 stellar values from \cite{Gillon:2017aa}. Using this set up we generate random uncertainties which are applied to all limb-averaged temporal outputs for all CO$_2$ cases. All spectra are binned in 10 pixel chunks to reach a precision $\lesssim$ 200 ppm in each spectral bin.

\begin{figure}[ht]
    \centering
    \includegraphics[width = 0.5 \textwidth]{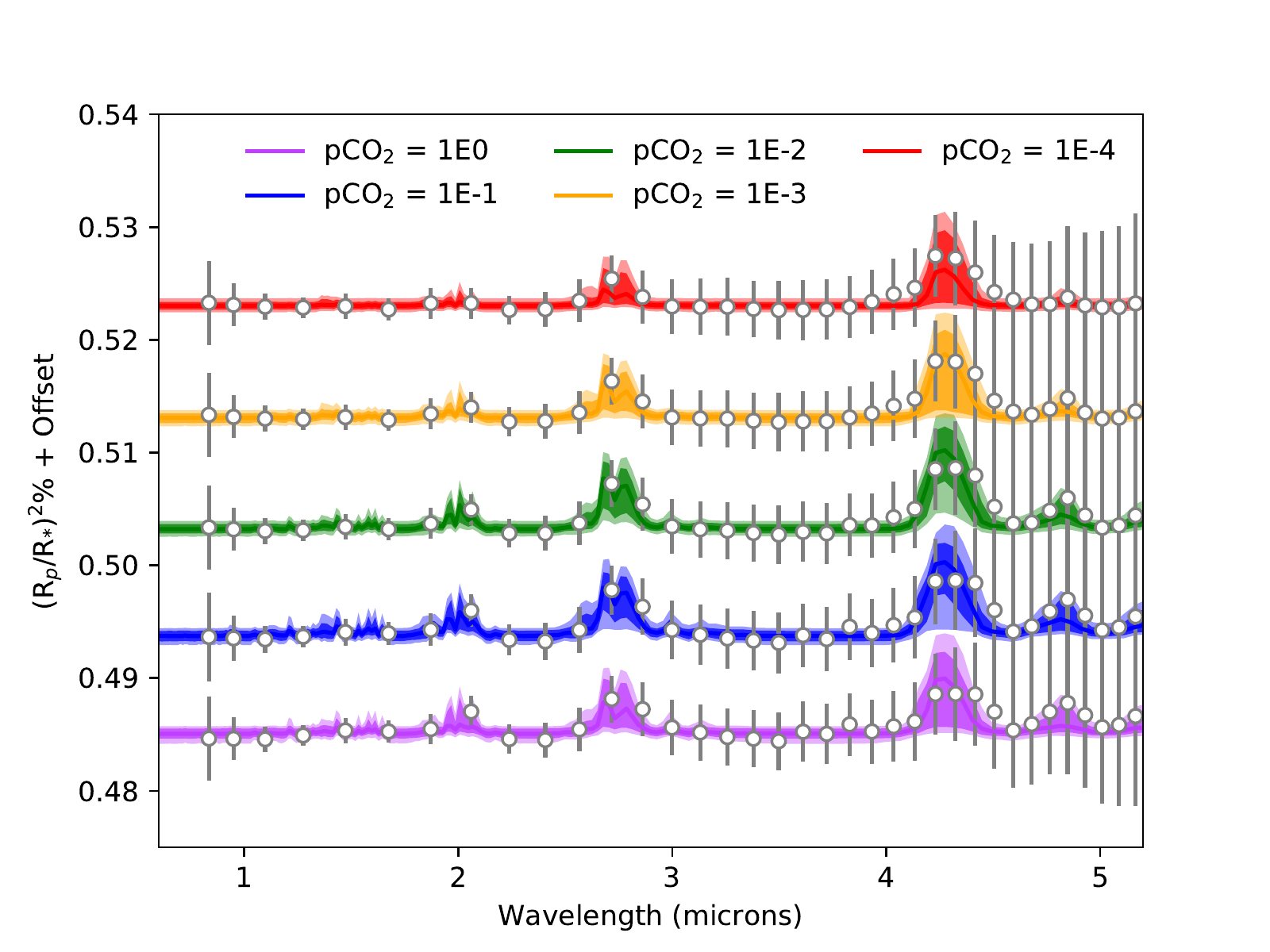}
    \caption{Sample retrieved spectra. Each line represents a different pCO$_2$ case for a single multi-epoch case. 1- and 2- $\sigma$ contours are included.}
    \label{fig:Retrieved_Spectra}
\end{figure}

As suggested in \cite{KrissansenTotton2018, Fauchez2019, LustigYaeger2019}, we assume that 10 transits are needed to achieve the S/N necessary to detect molecular features in the atmosphere of TRAPPIST-1e. To simulate the effect of these required multi-epoch observations, we randomly select 10 temporal output spectra and perform a weighted average of those times. This process is done 10 times, resulting in 10 multi-epoch `observations' for each CO$_2$ case. We use the {\texttt{PandExo}} uncertainties applied at the locations of the binned model, minimizing biases due to outliers in the random draws of the noise instance. \citep[e.g.,][]{Feng:2018aa,KrissansenTotton2018, Mai2019,Changeat2019, Taylor2020}. We generate the `observed' spectra before performing the weighted combination to capture the effect of the changing underlying model. The resulting data points and their corresponding uncertainties are shown in Figure \ref{fig:Retrieved_Spectra} for a single set of 10 combined epochs.

\section{Atmospheric Retrievals} \label{Retrievals}
To perform our retrievals we use the open source radiative transfer and retrieval framework  \texttt{CHIMERA} \citep{Line2013a,Line2013b}. It previously has been used to study the atmosphere of TRAPPIST-1e to determine the resolution that a future instrument needs to have in order to detect and constrain the abundances of molecules in the atmospheres of temperate terrestrial planets to high precision \citep{Tremblay:2020}. Our retrieval set up is similar to that presented in \citet{Tremblay:2020}. For each of the combined spectra we fit for 6 parameters: an isothermal temperature profile (T$_{\text{iso}}$), a radius scaling factor (x$R_{\text{p}}$), the cloud top pressure ($\log(\text{CTP})$), the mean molecular weight of the atmosphere and the volume mixing ratio of H$_2$O and CO$_2$. We use the mean molecular weight (MMW) as a proxy to effectively fill the rest of the atmosphere with N$_2$ as this molecule is inert. 
\section{Results} \label{Results}

\begin{figure*}[ht!]
    \centering
    \includegraphics[width = 0.8\textwidth]{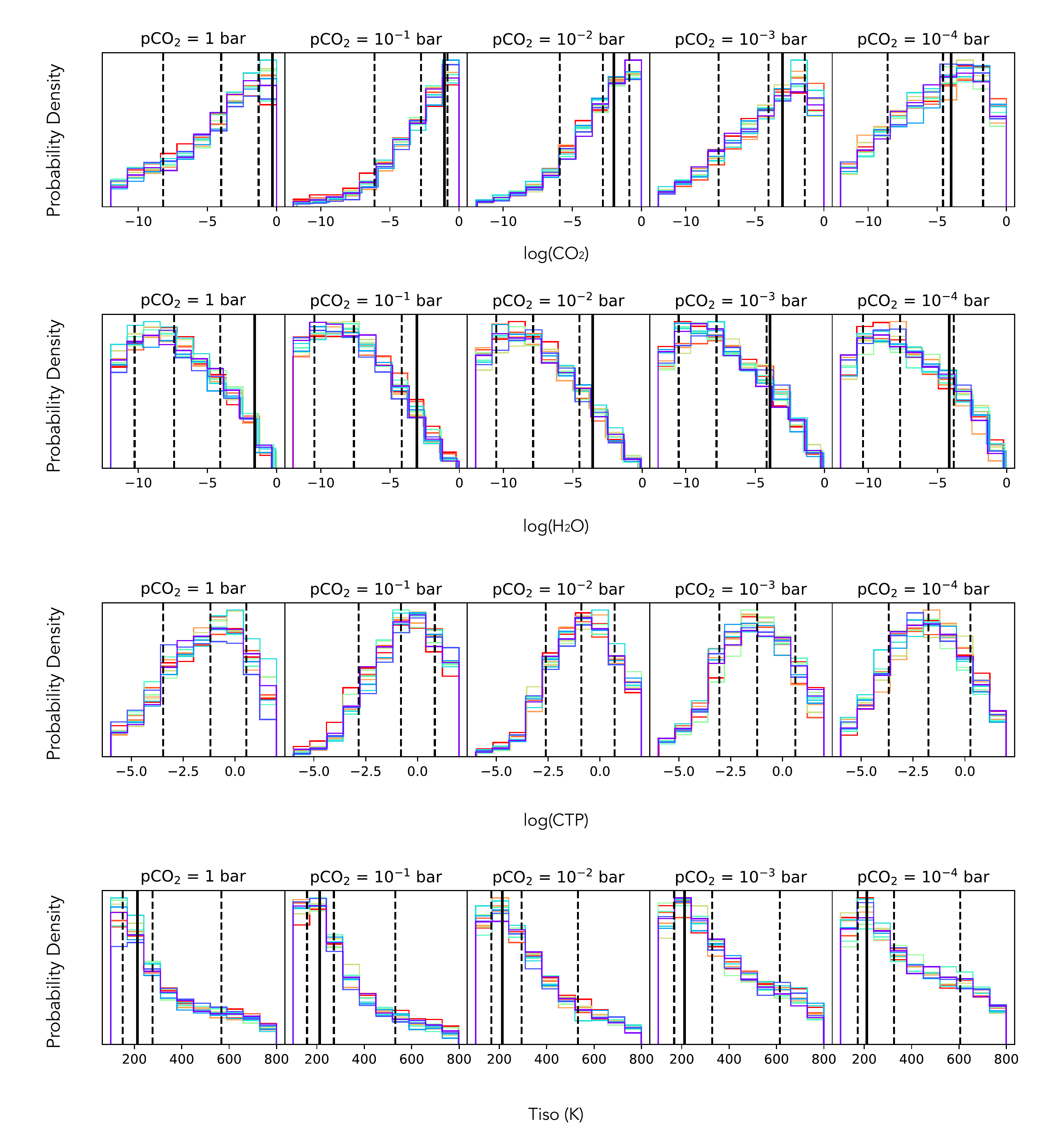}
    \caption{The retrieved posterior distributions for 4 of the key parameters explored in this study. From top to bottom the 4 panels are: CO$_2$ abundance, H$_2$O abundance, cloud top pressure and temperature.  Each coloured histogram represents the retrieved parameter for 1 of the combined spectra (representing 10 different ``observed'' epochs), for a total 10 different combined spectra, each with different underlying cloud assumption. The thick solid vertical lines show the terminator-average abundance of CO$_2$ and H$_2$O from the input models, while the dotted vertical lines represent the best fit value and the 1-$\sigma$ constraints on it. With the exception of the 1 bar pCO$_2$ case, the true CO$_2$ value is within 1-$\sigma$ of the retrieved values. We present the 0.1 bar temperature as a solid vertical line in the temperature histograms. It is not possible to present a "true" value for the cloud top pressure as this is different for each scenario. The consistency of the retrieved cloud top pressures demonstrates that the precision of the data is not high enough to detect this variability.}
    \label{fig:Histograms}
\end{figure*}

Each spectra that we retrieved on had a different underlying cloud configuration, hence we want to determine if these physical processes impact the retrieved CO$_2$ abundance. The histograms presented in Figure \ref{fig:Histograms} show that the retrieved abundances are consistent for each of the combined spectra, suggesting that the impact of cloud variability is not detectable with the resolution provided by \jwst\ NIRSpec PRISM. This is further verified by the 3rd panel which shows consistent retrieved cloud top pressure for each spectra. While the cloud top pressure can be seen to vary in our models, this consistency in the retrieved cloud top pressures demonstrates that the precision of \jwst\ NIRSpec PRISM will not be able to differentiate variable cloud spectra at any level of confidence. It can be seen from Figure \ref{fig:Histograms} that for the CO$_2$ partial pressure greater than 10$^{-1}$ bars the posterior tends towards the upper prior which is set by the physical limit of the volume mixing ratio, with the true value lying outside 1-$\sigma$. The lower partial pressures are not prior dominated and we are able to retrieve the correct volume mixing ratios within 1-$\sigma$. Further, the 10 cases for each partial pressure are consistent, suggesting that the variable clouds are not affecting the retrieved values. We are unable to constrain a water abundance due to its lack of strong spectral features in our models.

\section{Conclusions} \label{Conclusions}
We conducted GCM simulations that show inherent cloud variability along the limbs of tidally locked terrestrial planets, with specific results for TRAPPIST-1e. Variability in liquid water and water ice clouds is most pronounced near the limbs of the tidally locked planet, which critically lies in the region probed by transmission spectroscopy. The effect of this variability on simulated observed spectra is to change the spectral continuum as well as affect the shape of molecular absorption features.

While each resolution element in our 3D models experiences extreme variability in cloud coverage, the effect of intrinsic cloud variability on the limb averaged spectra is muted. For our simulated TRAPPIST-1e observations based on these 3D models, the precision of \jwst\ NIRSpec PRISM is not sufficient to detect the spectral variability due to the changing cloud cover, nor does it impact our ability to detect the atmosphere. Future work will explore a wider range of input scenarios to explore the limits of this effect, including its impact on next-generation observatories and other \jwst\ observing modes.

\acknowledgments
{E.M.M acknowledges support from APL's Independent Research and Development Program.

J.T. is a Penrose Graduate Scholar and would like to thank the Oxford Physics Endowment for Graduates (OXPEG) for funding this research.

T.D.K. acknowledges funding from the 51 Pegasi b Fellowship in Planetary Astronomy sponsored by the Heising–Simons Foundation. The GCM simulations were completed with resources provided by the University of Chicago Research Computing Center.

The entire team acknowledges the corgi Dexter, whose pictures got us through writing this manuscript.}

\software{\\ Astropy \citep{astropy,astropy2},
\\ IPython \citep{ipython},
\\ Matplotlib \citep{matplotlib},
\\ NumPy \citep{numpy, numpynew}
}

\bibliography{REFS}{}
\bibliographystyle{aasjournal}
\end{document}